\documentclass[10pt, a4paper, twocolumn, twoside]{article} 

\usepackage[english]{babel} 

\usepackage{microtype} 

\usepackage{amsmath,amsfonts,amsthm} 

\usepackage[svgnames]{xcolor} 

\usepackage[small, labelfont=bf, up]{caption} 

\usepackage{booktabs} 

\usepackage{lastpage} 

\usepackage{graphicx} 

\usepackage{enumitem} 
\setlist{noitemsep} 

\usepackage{sectsty} 
\allsectionsfont{\usefont{OT1}{phv}{b}{n}} 

\usepackage{geometry} 

\usepackage[T1]{fontenc} 
\usepackage[utf8]{inputenc} 

\usepackage{XCharter} 

\usepackage{hyperref}

\usepackage{journals}

\usepackage{fancyhdr} 
\newcommand{\shorttitle}[1]{\fancyhead[CE]{\textsl{#1}}}
\newcommand{\shortauthors}[1]{\fancyhead[CO]{\textsl{#1}}}

\fancypagestyle{firstpage}{ 
	\fancyhf{}
        \lhead{\vspace*{0.0em} \textit{\footnotesize 21$^{\rm st}$ European Workshop on White
            Dwarfs \\ Castanheira, Campos, Montgomery, eds. \\[-0.2em]
          July 23--27, 2018, Austin, Texas, USA}}
}

\date{}

\newcommand{\authorstyle}[1]{{\large\usefont{OT1}{phv}{b}{n}\color{DarkRed}#1}} 

\newcommand{\institution}[1]{{\footnotesize\usefont{OT1}{phv}{m}{sl}\color{Black}#1}} 

\usepackage{titling} 

\newcommand{\HorRule}{\color{DarkGoldenrod}\rule{\linewidth}{1pt}} 

\pretitle{
	\vspace{-30pt} 
	\HorRule\vspace{10pt} 
	\fontsize{16}{12}\usefont{OT1}{phv}{b}{n}\selectfont 
	\color{DarkRed} 
}

\posttitle{\par\vskip 10pt} 

\preauthor{} 

\postauthor{ 
	\vspace{0pt} 
	{\par\HorRule} 
	\vspace{-10pt} 
}

\newcommand{\newabstract}[1]{
    {\section*{Abstract}
    \bfseries #1}
  }

\usepackage[authoryear]{natbib}
\title{The Magnetic Fields of White Dwarfs in Cataclysmic Variables} 

\shorttitle{Magnetic fields in CVs} 

\shortauthors{Briggs et al.} 

\author{
        \authorstyle{Gordon~Briggs,$^1$ Lilia Ferrario,$^1$
          Christopher~Tout,$^{1,2}$ Dayal~Wickramasinghe$^1$}
	\newline\newline 
	$^1$\institution{Mathematical Sciences Institute, The
          Australian National University, ACT 2601, Australia; 
          Gordon.Briggs@anu.edu.edu}\\ 
	$^2$\institution{Institute of Astronomy, The Observatories, Madingley Road, Cambridge CB3 0HA; cat@ast.cam.ac.uk}\\ 
      }

\begin{document}

\maketitle 

\thispagestyle{firstpage} 


\newabstract{The origin of magnetic fields in isolated and binary
  white dwarfs has been investigated in a series of recent papers. One
  proposal is that magnetic fields are generated through an
  $\alpha$--$\Omega$ dynamo during common envelope evolution. Here we
  present population synthesis calculations showing that this
  hypothesis is supported by observations of magnetic binaries.}


\section{Introduction}

\begin{figure*}[htbp!]
\centering\includegraphics[width=0.8\linewidth]{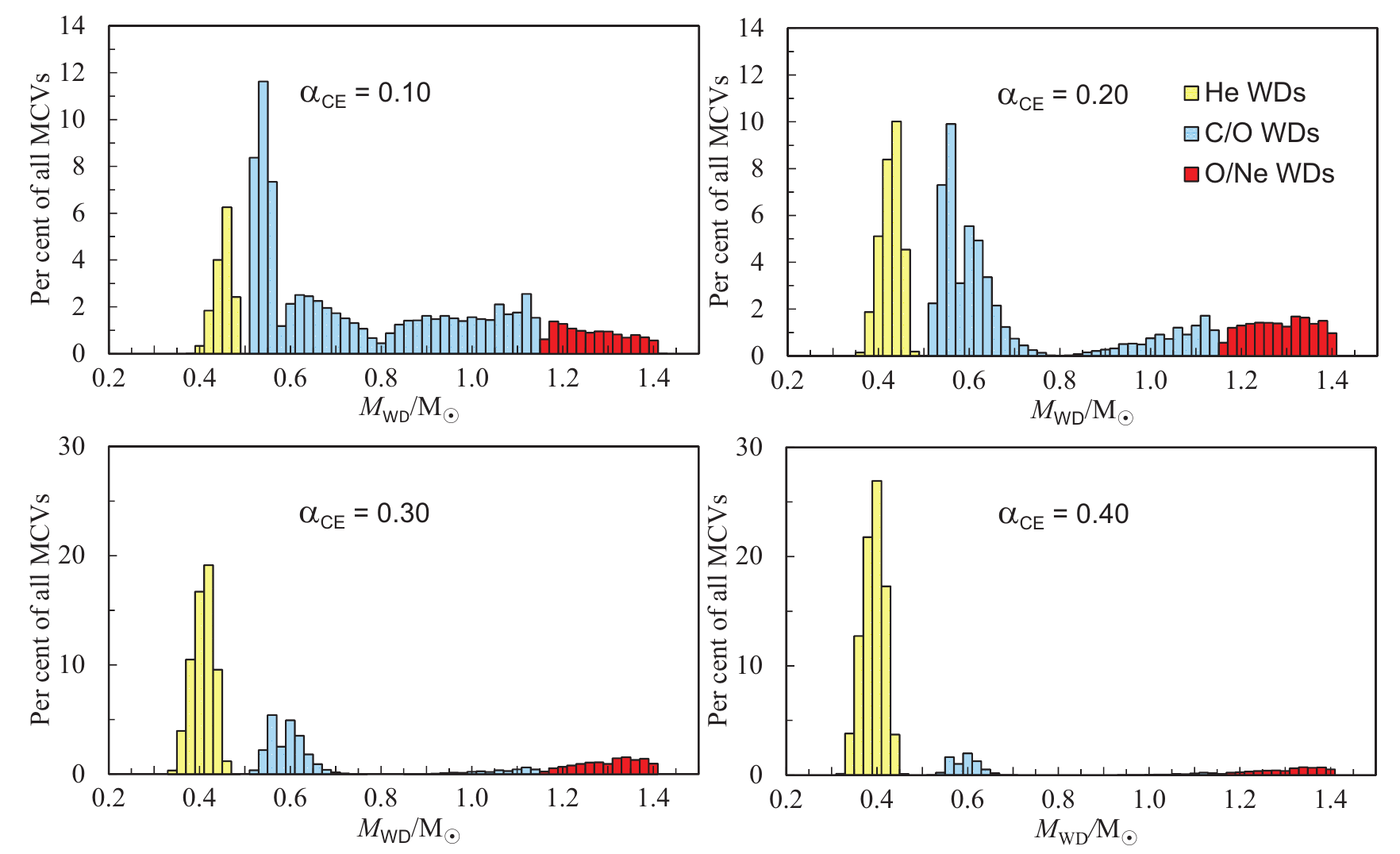}
\caption{Synthetic mass distribution of the magnetic WDs
  just before mass transfer begins.}
\label{Briggs_MassWDs}
\end{figure*}

The magnetic cataclysmic variables (MCVs) comprise a magnetic white
dwarf (MWD) accreting matter from a low-mass M-dwarf companion. The
accretion flows in MCVs are magnetically confined and form funnels in
the most strongly magnetic and synchronously rotating \emph{polars}
\citep{Ferrario1999} or curtains in the \emph{intermediate polars}
\citep[IPs,][]{FWK1993} where fields are not strong enough to prevent
the formation of a truncated accretion disk. The strength and
structure of the magnetic fields in polars have been established via
cyclotron and Zeeman spectroscopy \citep[][]{Ferrario1992,
  Ferrario1993STLMi,Ferrario1996,Schwope1999} to be in the range
$\sim 10^7$--$10^8$\,G \citep[see][and references
therein]{Ferrario2015MWD}. It is much more difficult to determine the
field strengths of the white dwarfs (WDs) in the IPs because the
radiation from these systems is dominated by their accretion disks
which swamp photospheric and cyclotron emission from the
accretion shocks \citep{FW1993}. Nonetheless, fields in IPs have been
estimated to be below a few $10^7$\,G
\citep[see][and references therein]{Ferrario2015MWD}.

In this paper we report population synthesis calculations that
investigate the hypothesis that fields in MCVs result from binary
interaction during common envelope (CE) evolution. This hypothesis was
first advanced by \citet{Tout2008}, following the earlier work of
\citet{Regos1995}, as an alternative to the fossil field hypothesis
\citep[e.g.,
see][]{Mestel1958,Woltjer1964,Angel1981,Tout2004,Ferrario2015Origin}.
According to \citet{Tout2008} the strong fields in MWDs derive from
the differential rotation generated by the stellar cores while they
spiral in toward each other during CE evolution. If they merge, they
give rise to a single, strongly magnetic WD. If they survive and
emerge from CE nearly in contact, they evolve into MCVs. The
calculations presented here for MCVs are an extension of those
performed by \citet{Briggs2015} and \citet{Briggs2018MWD} for the
isolated high-field magnetic WDs explained as the outcome of stellar
mergers \citep[see also][]{Nordhausetal2011, Garciaberro2012}.

\section{Calculations}

We have generated and evolved for $9.5\,$Gyr \citep[age of the
Galactic Disc,][]{Kilic2017} a synthetic population of binaries using
the rapid binary star evolution algorithm, {\sc bse}, of
\citet{Hurley2002}. The initial (main sequence) parameters are the
mass of the primary ($1.0-10.0$\,M$_\odot$), the mass of its companion
($0.1-2.0$\,M$_\odot$), and the orbital period
($1-10\,000$\,days). The mass of the primary follows Salpeter's mass
function while its companion is chosen to give a flat mass ratio $q$
distribution \citep{Hurley2002,Ferrario2012} with $q\le 1$.  The initial period
distribution is selected to be uniform in the logarithm.

  \citet{Briggs2018MWD} used the dynamo results of \citet{WTF2014} to
  assign a field to each of their synthetic WD resulting from stars that
  merge during CE evolution according to
\begin{equation}\label{Eqfield}
B = B_0\left(\frac{\Omega}{\Omega_{\rm crit}}\right)\, \mbox{G},
\end{equation}
\noindent where $\Omega$ is the orbital angular velocity of the system
at the point the envelope is ejected and $\Omega_{\rm crit}$ is the
break-up angular velocity of the nascent WD.  The parameter $B_0$ was
determined empirically by finding the best theoretical fit to the
observed field distribution of isolated MWDs
\citep{Briggs2018MWD}. While the shape and width of the field
distribution is determined by the CE efficiency parameter $\alpha$
\citep[found to be $\le$ 0.3, see][for further details on the
modelling procedure]{Briggs2015,Briggs2018MWD,Briggs2018MCV},
different $B_0$'s shift the field distribution to lower or higher
fields. Here we will show that the field prescription of equation
(\ref{Eqfield}) to model the fields of isolated MWDs can also
represent the field distribution of the MWDs in MCVs.
\begin{figure}[h!]
\includegraphics[width=\linewidth]{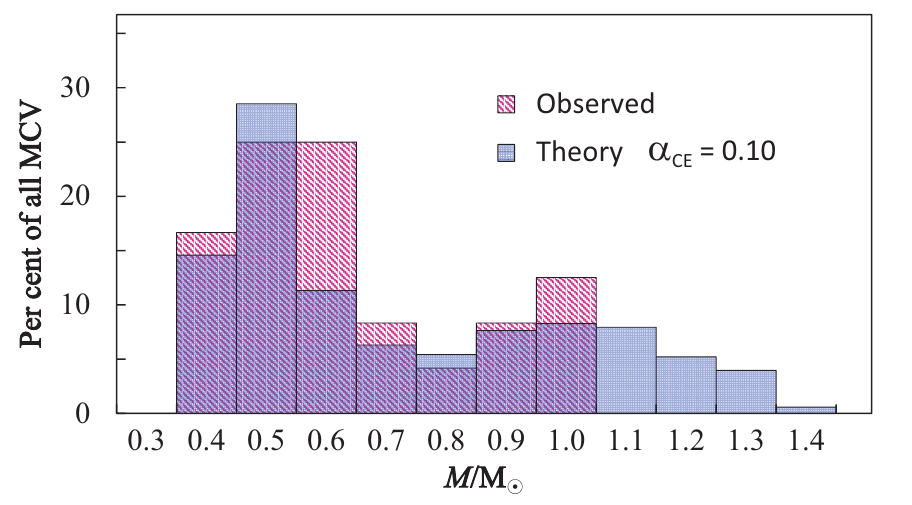}
\caption{Comparison of the theoretical mass distributions to observations
  from \citet{Zorotovic2011}.}
\label{Briggs_Mass_fit}
\end{figure}
\section{Results}

We have varied the CE efficiency parameter $\alpha$ to investigate its
effects on the theoretical population of MCVs.
Fig.\,\ref{Briggs_MassWDs} shows that at low $\alpha$ the type of
binaries that evolve to the beginning of Roche lobe overflow (RLOF)
mostly contain a CO\,WD.  As $\alpha$ increases the binaries with
He\,WDs become the predominant type. Because the observed fraction of
He\,WDs tends to be low among CVs \citep{Zorotovic2011} we can say
that a synthetic population produced with a low $\alpha$ can better
reproduce observations.

The WD mass distribution has a dip near $M_{\rm WD}=0.8$\,M$_\odot$
which widens as $\alpha$ gets larger.  This is caused by systems that,
as $\alpha$ increases, emerge from CE at longer orbital periods. But
the longer the period, the more massive the WD must be for mass
transfer to take place. Thus this gap is due to systems emerging from
CE at large separations but whose WDs are not enough massive to allow
RLOF to take place. The second narrower gap near 0.5\,M$_\odot$
separates systems with He\,WDs (CE occurred when the primary was an
RGB star) from those with CO\,WDs (CE occurred when the primary was an
AGB star). Because the current sample of PREPs
\citep[see][]{Ferrario2015MWD} is far too small, we have compared our
synthetic WD mass distribution to the observed sample of WDs in
non-magnetic Pre-CVs \citep{Zorotovic2011}. This comparison is shown
in Fig.\,\ref{Briggs_Mass_fit} for $\alpha = 0.10$ \citep[see][for
further details on the modelling procedure]{Briggs2018MCV}. We note
that the observed population exhibits the WD mass dip near
$0.8$\,M$_\odot$ that is predicted by theory. This comparison seems to
suggest that the WD mass distribution in pre-magnetic CVs does not
differ substantially from that in classical non-magnetic CVs
\citep[extreme cases such as PG1346+082][are not
considered in this context]{Provencal1997}.
\begin{figure*}[htbp!]
\centering\includegraphics[width=0.8\linewidth]{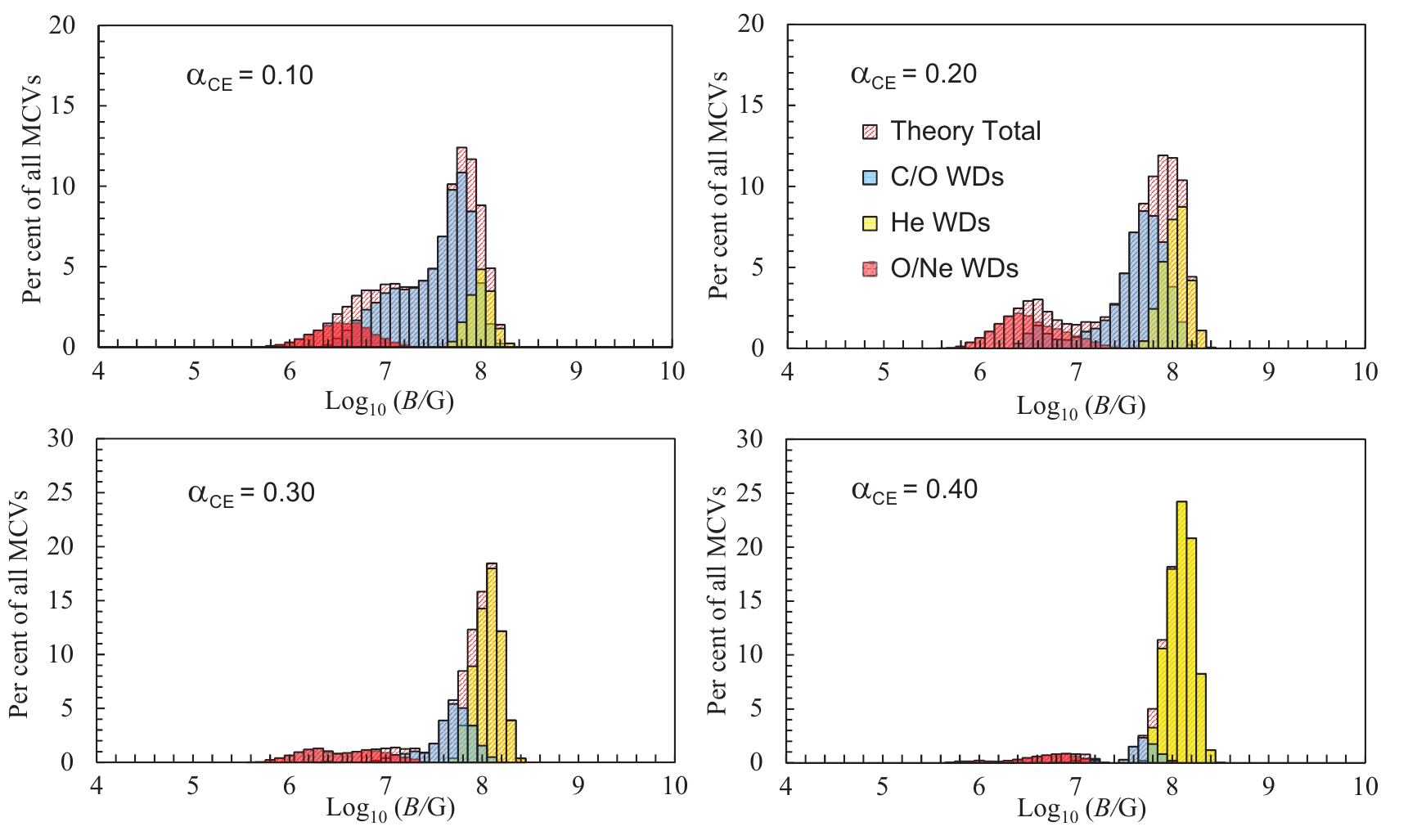}
\caption{Synthetic magnetic field distribution of the magnetic WDs
  just before mass transfer begins.}
\label{Briggs_Bmag}
\end{figure*}

We now analyse the field distribution of MCVs.
Fig.\,\ref{Briggs_Bmag} shows that the strongest magnetic systems are
those that host a He\,WD. The reason is that binaries that go through
CE evolution when the primary star is on the RGB have shorter orbital
periods and thus generate highly magnetic WDs, as expected from
equation (\ref{Eqfield}).  The field distribution is dominated by
binaries with CO\,WDs when $\alpha\le 0.2$. As $\alpha$ increases the
distribution becomes narrower, is dominated by He\,WDs, and moves to
higher fields. This is because most of those binaries that enter CE when the
primary is on the RGB end up merging at low $\alpha$ but do not at
high $\alpha$. Instead, they produce a population of strongly
magnetic, short period systems with low-mass He\,WDs.  The dip at
$8\times 10^6$\,G is caused by the dearth of WDs with masses near
0.8\,M$_\odot$ (see above) and the fact that equation (\ref{Eqfield})
links mass to field strength.

\begin{figure}[h!]
\includegraphics[width=\linewidth]{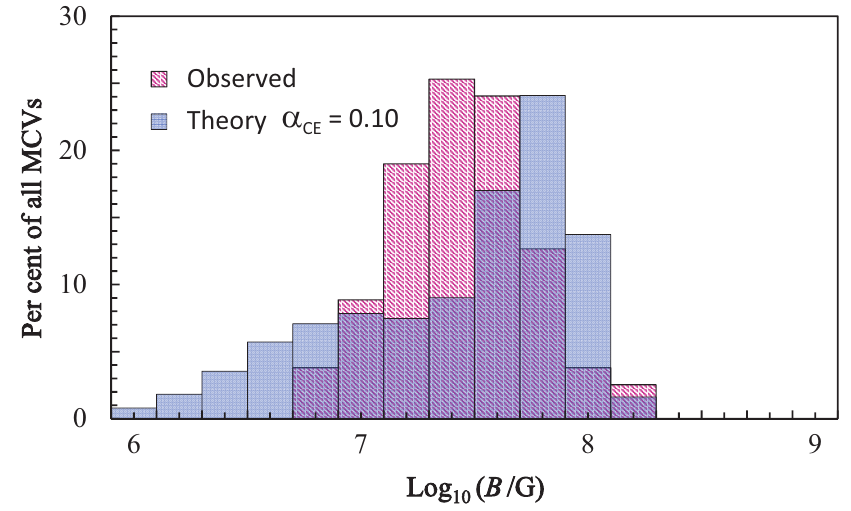}
\caption{Comparison of the theoretical field strength for
  $\alpha= 0.1$ and observations from \citet{Ferrario2015MWD}.}
\label{Briggs_Bmag_fit}
\end{figure}

The comparison between theoretical and observed field distributions is
shown in Fig.\,\ref{Briggs_Bmag_fit}. We note that we do not know what
the underlying (real) field distribution of the WDs in MCVs is at the
very low and very high ends of the field distribution. This is because
at low fields the observed radiation is dominated by the truncated
accretion disk so that photospheric Zeeman and cyclotron features are
not visible. These systems are thus excluded from the list of
\citet{Ferrario2015MWD} that only contains systems with field
measurements. At high fields the mass accretion from the companion
star is hindered \citep{LiWuWick1994,Li1998,Hoard2002,Hoard2004} so
that these systems are faint and thus difficult to detect.  Despite
these caveats the comparison of the magnetic field distribution shows
that the range of fields derived for our synthetic population is
consistent with observations of MCVs.

Despite the limitations of our modelling, we have illustrated that the
observed properties of the white dwarfs in MCVs can be explained in terms of a
population of binaries emerging from CE exchanging mass or close to
contact as first advanced by \citet{Tout2008}.  Our calculations are
also in support of the hypothesis that the low-accretion rate polars
(LARPS) are pre-MCVs \citep[LARPS were renamed PREPS by][to avoid
confusion with polars in a low state of accretion]{Schwope2009}.

The full results of these studies are reported in \citet{Briggs2018MCV}. 

\section*{Acknowledgements}

GPB gratefully acknowledges receipt of an Australian Postgraduate
Award.


\bibliography{Ferrario}

\end{document}